\documentclass[aps,prb,twocolumn,superscriptaddress,showpacs,floatfix]{revtex4}
\usepackage{graphicx}
\usepackage{latexsym}
\usepackage{stmaryrd}
\usepackage{amsmath}
\usepackage{amssymb}
\usepackage{epstopdf}
\usepackage[colorlinks,linkcolor=red,anchorcolor=red,citecolor=blue]{hyperref}
\DeclareMathOperator{\diag}{diag}

\begin{document}
\makeatletter
\newcommand{\rmnum}[1]{\romannumeral #1}
\newcommand{\Rmnum}[1]{\expandafter\@slowromancap\romannumeral #1@}
\newcommand{\B}[1]{{\textcolor{blue}{#1}}}
\makeatother

\title{Bulk-boundary correspondence in disordered higher-order topological insulators }

\author{Yu-Song Hu }
\affiliation{School of Physical Science and Technology, Soochow University, Suzhou, 215006, China}
\author{Yue-Ran Ding }
\affiliation{Institute for Advanced Study, Soochow University, Suzhou 215006, China}
\author{Jie Zhang }
\affiliation{School of Physical Science and Technology, Soochow University, Suzhou, 215006, China}
\author{Zhi-Qiang Zhang }\email{zhangzhiqiangphy@163.com}
\affiliation{School of Physical Science and Technology, Soochow University, Suzhou, 215006, China}
\author{Chui-Zhen Chen}\email{czchen@suda.edu.cn}
\affiliation{Institute for Advanced Study, Soochow University, Suzhou 215006, China}
\affiliation{School of Physical Science and Technology, Soochow University, Suzhou, 215006, China}
\date{\today}

\begin{abstract}
In this work, we study the disorder effects on the bulk-boundary correspondence of two-dimensional higher-order topological insulators (HOTIs). We concentrate on two cases: (i) bulk-corner correspondence, (ii) edge-corner correspondence. For the bulk-corner correspondence case, we demonstrate the existence of the mobility gaps and clarify the related topological invariant that characterizes the mobility gap.
Furthermore,  we find that,   while the system preserves the bulk-corner correspondence in the presence of disorder, the corner states are protected by the mobility gap instead of the bulk gap.
For the edge-corner correspondence case, we show that the bulk mobility gap and edge band gaps of HOTIs are no longer closed simultaneously.
Therefore, a rich phase diagram is obtained, including various disorder-induced phase transition processes.
Notably, a disorder-induced transition from the non-trivial to trivial phase is realized, distinguishing the HOTIs from the other topological states. Our results deepen the understanding of bulk-boundary correspondence and enrich the topological phase transitions of disordered HOTIs.
\end{abstract}

\maketitle

\section{Introduction}\label{section1}
The bulk-boundary correspondence is one of the most important features of the topological phase of matter.
For conventional TIs, the bulk-boundary correspondence is often referred to as bulk-edge correspondence indicating that $d$ dimensions systems have gapless states on the ($d-1$)-dimensional boundary.
Recently, the higher-order topological insulators (HOTIs) are introduced as a novel topological state
\cite{HOTI1,HOTI2,HOTI3,HOTI4,HOTI5,HOTI6,HOTI7,HOTI8,HOTI9,HOTI10,
HOTI11,HOTI12,HOTI13,HOTI14,HOTI15,HOTI16,HOTI17,HOTI18,HOTI19,
HOTI20,HOTI21,HOTI22,HOTI23,HOTI24,HOTI25,HOTI26,HOTI27,HOTI28,
HOTI29,HOTI30,BHZmodel,HOTI31,HOTI32,HOTI33,HOTI34,HOTI35,HOTI36,
HOTI37,HOTI38,HOTI39,HOTI40,HOTI41,HOTI42,HOTI43,HOTI44,HOTI45,
HOTI46,HOTI47,HOTI48,HOTI49,HOTI50,HOTI51,HOTI52,HOTI53,HOTI54,HOTI55}.
Generally, the HOTIs require the existence of both bulk energy gap and the non-trivial gaps of edge or surface states\cite{HOTI10,HOTI11} to ensure the existence of corner or hinge states.
Accordingly, the concept of the bulk-boundary correspondence for HOTIs is generalized to bulk-corner correspondence and edge-corner correspondence \cite{TI1,TI2,HOTIbe1,HOTIbe2}.
For example, for a two-dimensional HOTI\cite{HOTI10} constructed by the quantum spin Hall model\cite{TI1,TI2}, the helical edge states are gapped out with mass domains sitting at the edges of the sample, which guarantee the existence of the corner states and the non-trivial quadrupole moment \cite{HOTI1,HOTI2,HOTIqxy1,HOTIqxy2,HOTIqxy3}.
If the bulk energy gap and edge band gaps of HOTIs are closed simultaneously,
the HOTIs are considered to fall into the bulk-corner correspondence case \cite{TI1,TI2}.
On the other hand, when the edge band gap of HOTI is closed without bulk gap closing \cite{HOTIbe1,HOTIbe2},
the system belongs to the edge-corner correspondence instead of bulk-corner correspondence cases.

Meanwhile, the study on the disorder-induced phase transitions of higher-order topological phases also attracted great interest very recently\cite{HOTIq1,HOTIq2,HOTIq3,HOTIq4,HOTId1,HOTId4,HOTId5,HOTId6,HOTId7,HOTId8}.
The disorder-induced phase transitions in HOTIs\cite{HOTIq1,HOTIq2} show distinct features such as higher-order topological Anderson insulators (HOTAIs), comparing to those in conventional topological phases\cite{TAI0,TAI1,TAI2,TAI3,TAI4,TAI5,TAI6,TAI7,TAI8,TAI9}.
The numerical calculations indicate that the energy gap closes and reopens by the disorder, and the topological phase transitions are observed\cite{HOTIq1,HOTIq2}. In the study of Yang {\it et al}.\cite{HOTIq2}, the existence of extended bulk states is discussed. Furthermore, the presence of mobility edge is also clarified.
 In Li {\it et al}.'s study, however, they pay more attention to the edge band gaps. The edge band gaps are considered to be closed and reopened\cite{HOTIq1}. They also notice that the localization length along a specific direction diverges for samples under open boundary conditions, similar to those in one-dimensional Su-Schrieffer-Heeger models\cite{TAI0}.

However, previous studies mainly concentrated on the evolution of the bulk energy gap and edge energy gap. Although the mobility gap for HOTIs has been discussed, the topological invariant protected by the mobility gap is still unclear\cite{TAI6,TAI7,TAI8,TAI9}.
Furthermore, to demonstrate the influence of disorder on the bulk-corner correspondence, it is of great value to study the evolutions of mobility gaps and the correlated topological invariant.
More importantly, due to the breakdown of bulk-corner correspondence,
the disorder-induced gap closing and the related topological phase transitions of HOTI with edge-corner correspondence should be abundant.
 It is essential to study the disorder-induced gap closing of HOTIs, which will deepen the understanding of the topological features of the disordered HOTIs.

In this paper, we study the disorder effects on the evolution of both edge and bulk gaps, as well as the bulk-boundary correspondence in HOTIs. We investigate both the bulk-corner correspondence and edge-corner correspondence cases. For the bulk-corner correspondence case, we first demonstrate the existence of mobility gaps. Further, the topological invariant named as spin Chern number\cite{CS1,CS2,CS3} $\mathcal{C}_s$ is revealed to characterize the bulk states. By comparing the spin Chern number with the quadrupole moment, we find bulk-corner correspondence of the HOTIs remains intact in the presence of disorder. A phase diagram is given to elucidate the evolution of the mobility gap by varying the disorder strength.
For the edge-corner correspondence case, the disorder-induced edge energy gap closing is achieved. Importantly, unlike the previously discovered HOTAI, a phase transition from topological non-trivial phases to trivial ones is identified due to the competition between two renormalized parameters.
In addition, a variety of gap-closing processes with disorder strength $W\leq5t$ (where the band renormalization dominates) are obtained, which enriches the disorder-induced phase transitions of HOTIs.

The rest of this paper is organized as follows: In Sec.~\ref{section2}, we present the details of the model and the methods. In Sec.~\ref{section3}, we concentrate on the gap closing and phase transitions of the bulk-corner correspondence cases. In Sec.~\ref{section4}, The edge-corner correspondence cases are considered. Finally, a brief discussion and summary
are presented in Sec.~\ref{section5}.

\section{model and methods}\label{section2}

\begin{figure}[t]
   \centering
    \includegraphics[width=0.45\textwidth]{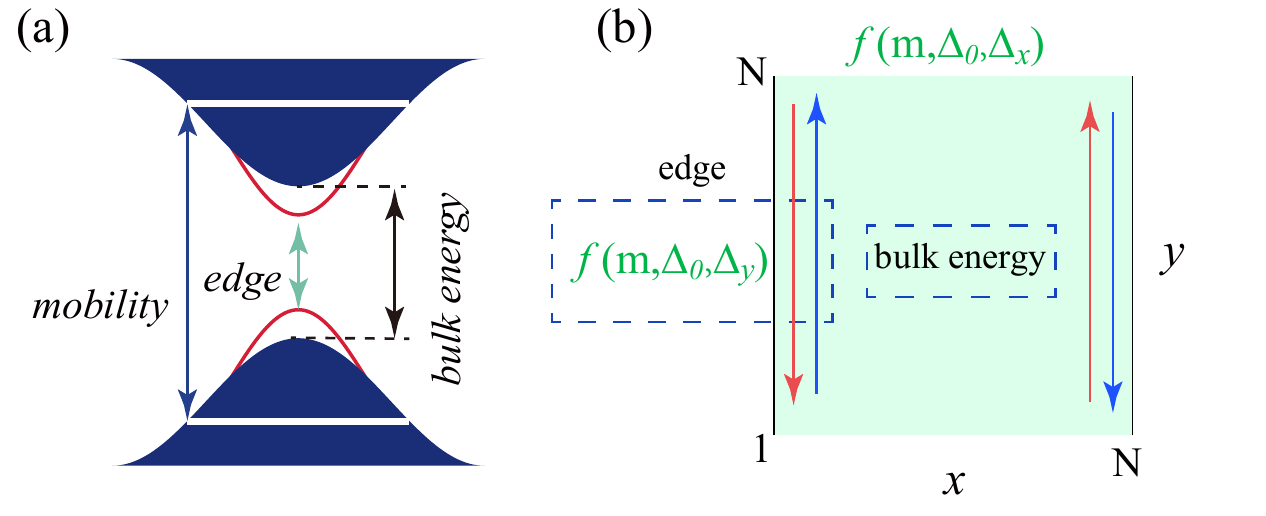}
    \caption{(Color online). Three different gaps for higher-order topological insulators, which are edge band gap, bulk energy gap, and mobility gap. (b) The schematic diagram of sample in real space with $N\times N$ primitive cells. For clean samples, the edge band gap along $x$ and $y$ directions are $f(m,\Delta_0,\Delta_x)$ and $f(m,\Delta_0,\Delta_y)$, respectively. $f(m,\Delta_0,\Delta_{x/y})$ is a function of $m$, $\Delta_0$ and $\Delta_{x/y}$. When $f(m,\Delta_0,\Delta_x)>0$ and $f(m,\Delta_0,\Delta_y)<0$, the edge band gaps construct the mass domain walls. The red and blue arrows are the helical edge states. We concentrate on the edge band gap along $y$ direction.}
   \label{f1}
\end{figure}

Following previous studies, the Hamiltonian can be written as \cite{HOTI30,HOTIbe1}:
\begin{align}
\begin{split}
\mathcal{H}(\textbf{k})&= [-m+t\cos k_x+t\cos k_y]\tau_z\sigma_z\\
&+\lambda \sin k_x\tau_x\sigma_z-\lambda \sin k_y\tau_y\sigma_z\\
&+\Delta(k_x,k_y)\tau_0\sigma_x,
\label{EQ1}
\end{split}
\end{align}
in the basis $(A_\uparrow,A_\downarrow,B_\uparrow,B_\downarrow)$ with $A,B$ ($\uparrow,\downarrow$) representing the orbit (spin) degrees of freedom. The Pauli matrices (identity matrix) $\tau_{x/y/z}$ ($\tau_0$) and $\sigma_{x/y/z}$ ($\sigma_0$) act on orbit and spin space, respectively.
$m$, $\lambda$ and $t$ are model parameters and we set $\lambda=t$ throughout the paper.
Here $\Delta(k_x,k_y)=\Delta_0+\Delta_x\cos k_x+\Delta_y\cos k_y$ with $\Delta_x=-\Delta_y$.
If $\Delta(k_x,k_y) = 0$, the Hamiltonian is exactly the Bernevig-Hughes-Zhang (BHZ) model\cite{BHZmodel}
and can describe a quantum spin-Hall effect with a pair of helical edge modes.
Generally, when $\Delta(k_x,k_y)\neq 0$ , the edge helical states with opposite group velocities will couple
and gapped out [see Fig.~\ref{f1}(a)].
To be specific, the $(\Delta_x\cos k_x-\Delta_x\cos k_y)$ can create the mass domain walls along the edge of the sample and lead to a HOTI with corner states.
Importantly, the bulk-corner correspondence exists only when $\Delta_0=0$
 and  the Hamiltonian has the edge-corner correspondence instead  for $\Delta_0\neq 0$ \cite{HOTIbe1} [see appendix for more details].

Before starting calculation, we define the quadrupole moment\cite{HOTIq1,HOTIq2}:
\begin{equation}
Q_{xy}=\frac{1}{2\pi}Im\{log[det(U^\dagger \widehat{q}U)\sqrt{det(\widehat{q}^\dagger)}]\},
\label{EQ2}
\end{equation}
 where $\widehat{q}\equiv exp[i2\pi \widehat{x}\widehat{y}/(N^2)]$.
 $N$ is the sample size and $U$ is constructed by the eigenvalues of occupied states with the projection operator $P_r=UU^\dagger$. $\widehat{x}$ ($\widehat{y}$) is the coordinate operator along $x$ ($y$) direction.
Note that $Q_{xy}$ is quantized since the Hamiltonian $\mathcal{H}$ has the chiral symmetry $\mathcal{S}=\tau_0\sigma_y$, i.e. $\mathcal{S}\mathcal{H}(\textbf{k})\mathcal{S}^{-1}=-\mathcal{H}(\textbf{k})$.

In addition to the quadrupole moment $Q_{xy}$, we will use the spin Chern number  $\mathcal{C}_{s}=|\mathcal{C}_+-\mathcal{C}_-|/2$ \cite{CS1,CS2,CS3}. Here the Chern number $\mathcal{C}_{\pm}$ for different spin is given by\cite{HOTId4,HOTId7,CS1}:
\begin{equation}
\mathcal{C}_{\pm}=-\frac{2\pi i}{N_xN_y}\sum_{n,\alpha}\langle n,\alpha|P_{\pm}[-i[\widehat{x},P_{\pm}],-i[\widehat{y},P_{\pm}]] |n,\alpha\rangle,
\label{EQ3}
\end{equation}
 $|n,\alpha\rangle$ is the eigenvector for site $n$ and orbit $\alpha$.
 $P_{\pm}=U_{\pm}U_\pm^\dagger$ are the projection operators for different spin components, and $U_{\pm}$ are determined from the eigenvectors of $P_r(\tau_0\sigma_z) P_r$.
 Although the spin up ($\uparrow$) and spin down ($\downarrow$) parts of the Hamiltonian are coupled by $\Delta(k_x,k_y)$, the spin Chern number is still well defined when the eigenvalue of $P_r(\tau_0\sigma_z) P_r$ has a gap\cite{CS1}.  $C_s=1$ suggests the existence of edge states and the non-trivial features of bulk states.

At last, the Anderson disorder is introduced as \cite{HOTId7} $H_n^w= \diag\{\varepsilon_{n}^{1},\varepsilon_{n}^{2}\} \otimes\sigma_z$, to
preserve the symmetry $\mathcal{S}H^w_n\mathcal{S}^{-1}=-H_n^w$. 
 The random onsite potential $ \varepsilon_{n}^{1}$, $ \varepsilon_{n}^{2}$ for site $n$ satisfy the uniform distribution $\varepsilon_{n}^{1},\varepsilon_{n}^{2}\in[-W/2,W/2]$ with the disorder strength $W$.
For a weak disorder strength, the renormalization of model parameters can be evaluated via the self-consistent Born approximation (SCBA) of self-energy \cite{HOTId7,HOTIq1,HOTIq2}
\begin{equation}
\Sigma^r=\sum_{i=1}^{2}\frac{W^2}{48\pi^2}\iint_{BZ}d^2\textbf{k}\gamma_i[E_F+i0^+-
\mathcal{H}(\textbf{k})-\Sigma^r]^{-1}\gamma_i.
\label{EQ4}
\end{equation}
with $\gamma_{1,2}=(\tau_0 \pm \tau_z)\otimes\sigma_z$/2. The integral is on the first Brillouin zone
and $E_F$ is the Fermi energy.

\begin{figure}[b]
   \centering
    \includegraphics[width=0.45\textwidth]{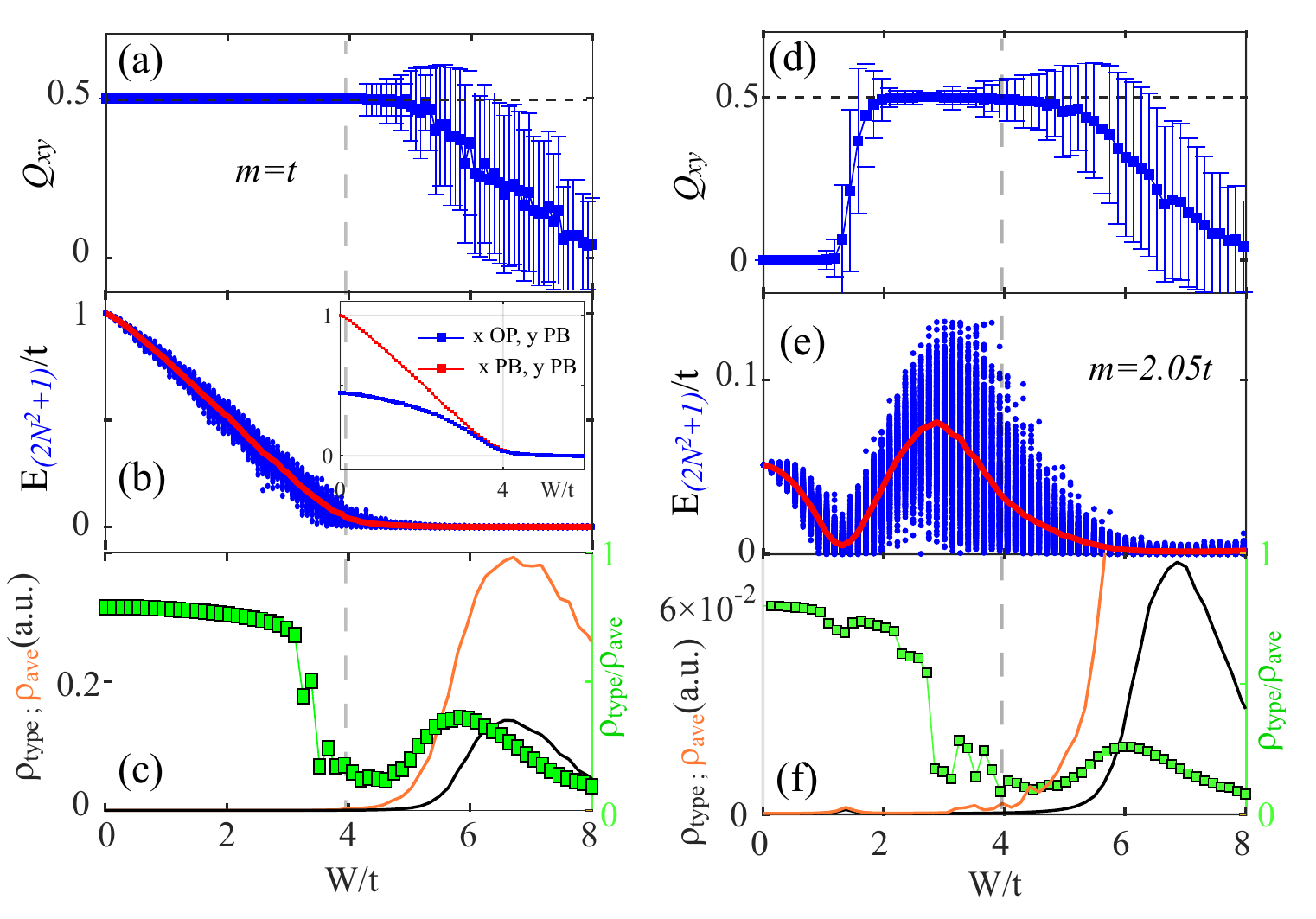}
    \caption{(Color online). (a), (b), and (c) are the quadrupole moment $Q_{xy}$, the eigenvalue $E_{2N^2+1}$ and DOS versus disorder strength $W$, respectively. The parameters are set as $m=t$, $\Delta_x=-\Delta_y=0.5$, $\Delta_0=0$. In our numerical calculations, a square sample with size $N$ is used. It has $4N^2$ eigenvalues in total. The bulk gap equals to $2E_{2N^2+1}$. The average DOS and the geometric average DOS are  $\rho_{ave}$ and $\rho_{type}$, respectively. The color of the curves are the same as the labels for (c) and (f). The blue (red) line for inset of (b) shows $E_{2N^2+1}$ versus $W$ for open (periodic) boundary condition, which shows the evolution of bulk (edge) gap. OB and PB corresponds to the open and periodic boundary condition, respectively. (d)-(f) are the same with (a)-(c), except that $m=2.05t$. The dashed lines roughly shows the bulk energy gap closing points. The sample size is $N=40$. }
   \label{f2}
\end{figure}

\section{HOTI with bulk-corner correspondence}\label{section3}

Different from the traditional two-dimensional topological insulators, the HOTIs could have the bulk-corner correspondence or edge-corner correspondence\cite{HOTIbe1}.
In this section, we consider $\Delta_0=0$ in Eq. (\ref{EQ1}) and the Hamiltonian of the HOTI has the bulk-corner correspondence.

\subsection{ mobility gap and the related topological invariant}

For the clean HOTIs with bulk-corner correspondence, the evolutions of the bulk energy gap are correlated with the edge band gaps and the topological phase transitions.
To be specific, the bulk energy gap closing determines the topological phase transition of the HOTIs in the same way as in the traditional topological insulators.
However,  when the disorder is considered, the bulk gap will be replaced by the mobility gap for both the 2D traditional topological insulators\cite{TAI6,TAI7,TAI8,TAI9} and HOTIs\cite{HOTIq1}.

To demonstrate the existence of the mobility gap of HOTI, we compare the $Q_{xy}$
with both the bulk energy gap and the edge energy gap and plot them as a function of the disorder strength $W$ in Fig.~\ref{f2}.
The $Q_{xy}$  and the bulk energy gap are evaluated utilizing a square sample with size $N$ [see Fig. \ref{f1}(b)] under periodic boundary conditions in both $x$ and $y$ directions.
On the other hand, the edge energy gap is calculated with open and periodic boundary conditions in $x$ and $y$ directions, respectively.
Because the whole spectrum has $4N^2$ eigenvalues in total as well as symmetry $\mathcal{S}$, The energy gap is determined by the $(2N^2+1)$-{\it th} eigenvalue $E_{2N^2+1}$ and the energy gap closes when $E_{2N^2+1}=0$.

In Figs.~\ref{f2}(a) and (d), the system starts from a HOTI phase, and a normal insulator exists in the clean limit. For a weak disorder, one can see a disorder-induced topological phase transition from a normal insulator to a HOTI phase in Fig.~\ref{f2} (d) and the bulk energy gap closes in Fig.~\ref{f2} (e), which correspond to the creation of the HOTAI.
When the disorder is strong enough, bulk energy gaps are closed for both HOTIs and HOTAIs in Figs~.\ref{f2}(b) and (e).
Notably, the $Q_{xy}$ is still quantized when the bulk energy gap closes at $W\approx 4t$. This means that the topological phase transition from HOTI to normal insulator is independent of the bulk energy gap when disorder $W\geq4t$.

In order to clarify the localization properties of bulk states, we calculate the average density of states (DOS) and the geometric average DOS\cite{TAI6,TAI7,TAI8} given by
\begin{align}
\begin{split}
\rho_{ave}(E)=\sum_{i=1}^L\rho_i/L,~~\rho_{type}(E)=\exp[\sum_{i=1}^L\ln \rho_i/L].
\end{split}
\end{align}
where $L$ is the number of the considered sites.
$\rho_i(E)=-\frac{1}{\pi}Im[G^r_{i,i}(E)]$ with $G^r(E)=[E+i0^+-H]^{-1}$, and $H$ is the real space Hamiltonian.
In Figs.~\ref{f2}(c) and (f), $\rho_{ave}$, $\rho_{type}$ and the ratio $\rho_{type}/\rho_{ave}$ are also plotted as a function of $W$.
Generally, $\rho_{type}$ is smaller than $\rho_{ave}$, $\rho_{type}/\rho_{ave}$ tends to be finite\cite{TAI6,TAI7,TAI8} for extended states and $\rho_{type}/\rho_{ave}$ will approach zero when the states are much more localized.
In Fig.~\ref{f2}(c) when $W>4t$, the density of state $\rho_{ave}\neq0$ meaning that the bulk energy gap disappears. In this circumstance, the quantized quadrupole moment $Q_{xy}=1/2$ is protected by the mobility gap instead.
By further increasing disorder strength to $W\approx 6t$, $\rho_{type}/\rho_{ave}$ shows a peak and $Q_{xy}$ decreases to about $1/4$ [half of its quantized value], meaning that the mobility gap is closed with the emergence of delocalized states at $E=0$.

\begin{figure}[t]
   \centering
    \includegraphics[width=0.45\textwidth]{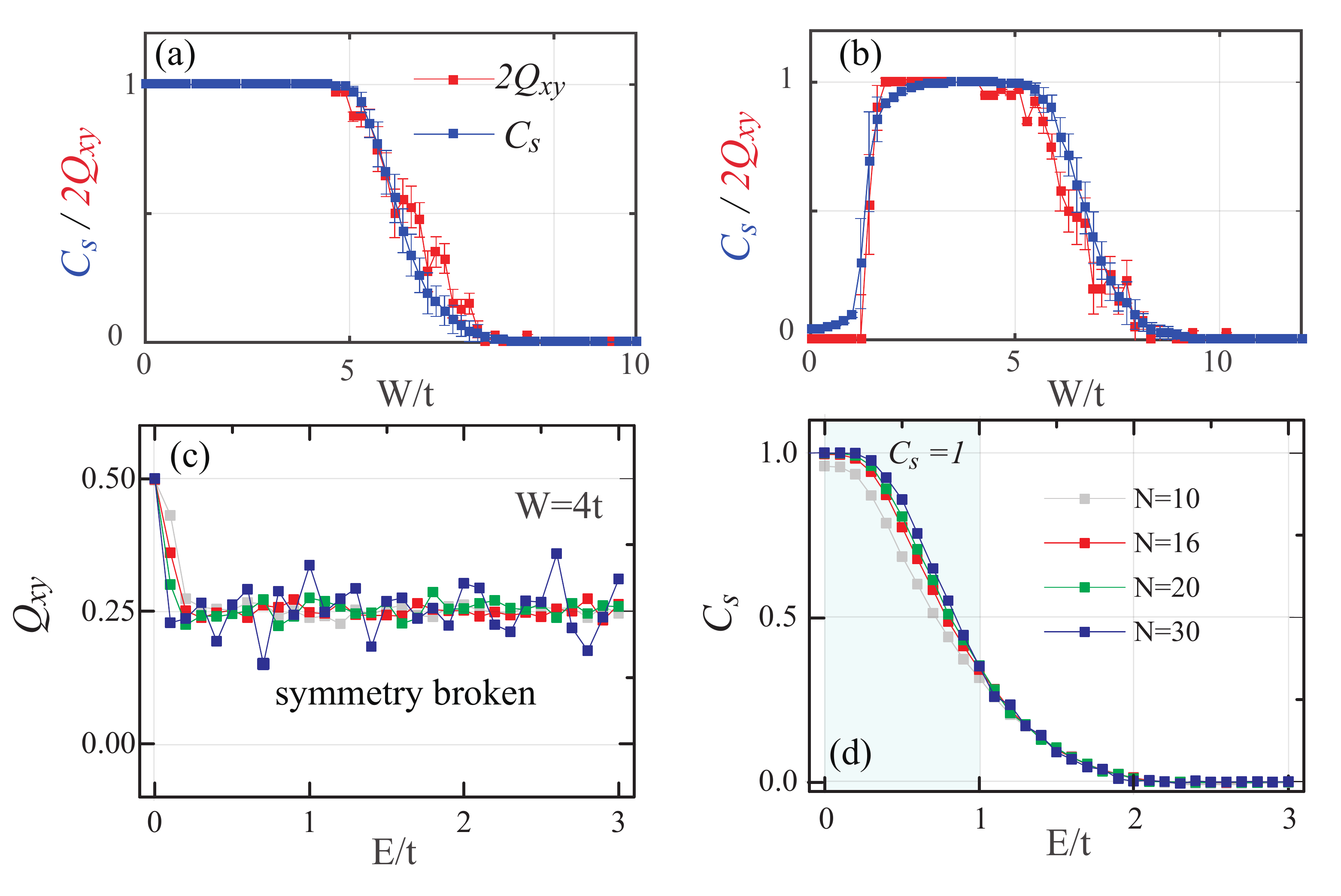}
    \caption{(Color online). The spin Chern number $C_s$ and quadrupole moment $Q_{xy}$ versus disorder strength $W$. (a) $m=t$, (b) $m=2.05t$. The sample size is $N=40$. (c) and (d) are the scaling of $Q_{xy}$ and $C_s$, respectively. Disorder strength is fixed at $W=4t$ with $m=t$. Other parameters are the same as those in Fig. \ref{f2}. }
   \label{f3}
\end{figure}

\begin{figure*}[t]
   \centering
    \includegraphics[width=0.8\textwidth]{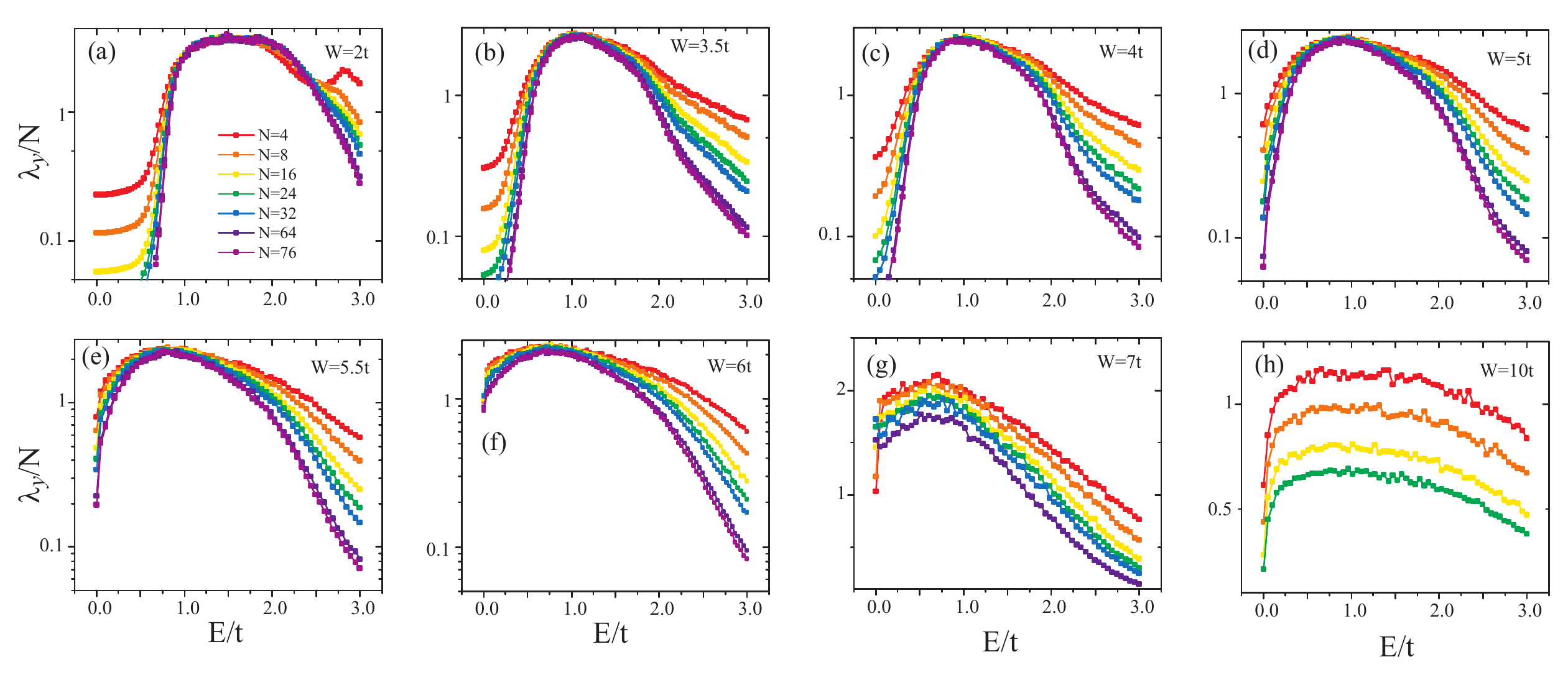}
    \caption{(Color online). The normalized localization length $\lambda_y/N$ versus energy $E$ for different sample sizes $N$. The disorder strength is: (a) $W=2t$, (b) $W=3.5t$, (c) $W=4t$, (d) $W=5t$, (e) $W=5.5t$, (f) $W=6t$, (g) $W=7t$, (h) $W=10t$, respectively. Parameters are set as $m=t$, $\Delta_0=0$, and $\Delta_x=0.5t$. The periodic boundary condition is adopted for (a)-(h). }
   \label{f4}
\end{figure*}

In two dimensions, all the states tend to be localized in the presence of disorder, and the emergence of delocalized states is generally accompanied by topological invariants \cite{MBL1,MBL2}.
Therefore, in the following, we explore the topological invariants that characterize the topological phase in the mobility gap.
To start with, we plot the quadrupole moment $Q_{xy}$ as shown in Fig. \ref{f3}(c).
Unfortunately, we find that the quantization of $Q_{xy}=0.5$ is destroyed when the Fermi energy slightly deviates from $E=0$. That's because the quadrupole moment $Q_{xy}$ is protected by the symmetry $\mathcal{S}=\tau_0\sigma_y$, which is preserved only when the system is half-filled, namely $E=0$.

Next, we consider the spin Chern number $\mathcal{C}_s=1$ to distinguish the HOTI from the trivial ones,
because Eq.~(\ref{EQ1}) can be viewed as the quantum spin Hall state with gapped edge states and the bulk-edge correspondence also holds (see appendix).
Different from the Chern number, $\mathcal{C}_s=1$ can determine the existence of gapped edge states, which are essential for HOTIs.
As illustrated in Figs.~\ref{f3}(a) and (b), the spin Chern number $\mathcal{C}_s$ perfectly fits the evolution of $Q_{xy}$ by increasing disorder strength. The quadrupole moment equals to $0.5$ when $\mathcal{C}_s=1$. Besides, $\mathcal{C}_s$ also captures the existence of HOTAI, shown in Fig. \ref{f3}(b). Thus, it is appropriate to adopt the spin Chern number to determine HOTIs within the mobility gap. Regarding $\mathcal{S}\mathcal{H}\mathcal{S}^{-1}=-\mathcal{H}$,
the system is symmetric about $E=0$, and we only pay attention to $E\geq0$ in all our calculations.
In Fig.~\ref{f3}(d), we do finite-size scaling of $\mathcal{C}_s$  and find
$\mathcal{C}_s\rightarrow 1$ for energy roughly within $E\in[0,t]$.
This confirms the quantization of $C_s$ within the entire mobility gap of the disordered HOTIs.

Generally, the $C_s=1$ only ensures the existence of gapped edge states within the mobility gap.
Nevertheless, due to the overlaps between $Q_{xy}$  and $\mathcal{C}_s$ curves [see Figs. \ref{f3}(a) and (b)],
it is appropriate to conclude that $\mathcal{C}_s=1$
accompanied with $Q_{xy}=0.5$ at $E=0$ should also ensure the existence of corner states in the mobility gap.
This interprets why the HOTIs could have the bulk-corner correspondence.

\subsection{mobility gap closing and phase transition}

In order to study the evolution of the mobility gap versus disorder, we calculate the normalized localization length\cite{NL0,NL1,NL2,NL3,NL4,NL5} $\lambda_y/N$ along $y$ direction with $m=t$.
The localized and delocalized states can be distinguished by the finite-size scaling analysis of $\lambda_y/N$.
 $\lambda_y/N$ decreases (increases) for localized (extended) states with $N$ increasing, while $\lambda_y/N$ is independent of $N$ at critical points.
For small disorder strength $W$ in Figs.~\ref{f4}(a)-(c), the system shows a mobility gap near $E=0$ with $\lambda_y/N$ decreasing, and then it encounters a critical point.  Notably, for $W=4t$, the critical point is at $E\approx t$, which is consistent with the scaling of $\mathcal{C}_s$ in Fig. \ref{f3}(d).

With an increase of $W$, the mobility gap decreases and eventually disappears when $W\approx 7t$ [see Figs.~\ref{f4}(d)-(h)],
 where all the states are localized for the entire energy range. Note that similar results can also be found for HOTAI with $m=2.05t$, as shown in the appendix.

\begin{figure}[t]
   \centering
    \includegraphics[width=0.49\textwidth]{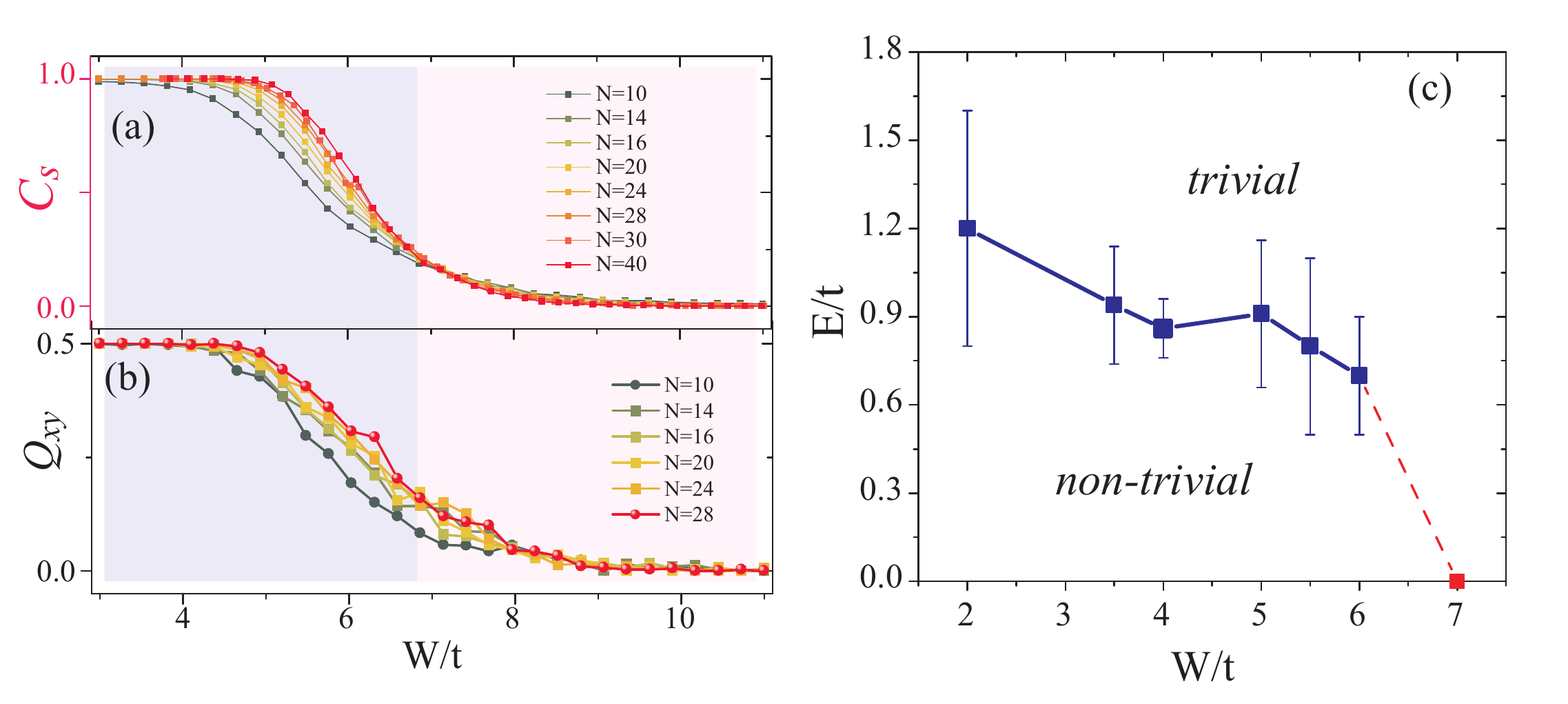}
    \caption{(Color online). (a) the scaling of $C_s$ versus $W$ with $E=0$. (b) is the same with (a), expect for $Q_{xy}$ instead. (c) The plot of mobility edge versus disorder strength $W$. The blue squares are obtained with the help of localization length. The red dashed line and the red square are determined by (a). $m=t$, $\Delta_x=0.5t$, and $\Delta_0=0$ for all the plots.   }
   \label{f5}
\end{figure}

To further demonstrate that the HOTI becomes a trivial insulator when the mobility gap closes, we perform the scaling analysis of the spin Chern number and quadrupole moment under different disorder strengths. As shown in Figs.~\ref{f5}(a) and (b), $\mathcal{C}_s\rightarrow 1$ in the thermodynamic limit when $W<6.8t$, while $Q_{xy}$ increases monotonously with the increase of $N$ in this disorder region.
On the other hand, when $W>6.8t$, both $Q_{xy}$ and $\mathcal{C}_s$ decrease with increasing $N$, and they tend to approach zero for larger $N$. These results are in accordance with those in Fig. \ref{f3}(a), where the $\mathcal{C}_s$ and $Q_{xy}$ curves almost overlap.

We summarize the above results in the phase diagram in Fig.~\ref{f5}(c). The blue squares are obtained by scaling of the normalized localization length in Fig.~\ref{f4}, while the red square is obtained with the help of Fig. \ref{f5}(a) and (b). One can see the HOTAIs (with $Q_{xy}=0.5$) and the trivial insulator phase (with $Q_{xy}=0.0$) are separated by the mobility edge.
Therefore, we conclude that the HOTIs or HOTAIs are protected by the mobility gap in the presence of disorder.

We close this section by discussing the relation between the edge band gap and the existence of corner states.
The inset of Fig. \ref{f2}(b) shows the edge band gap and bulk band gap are closed simultaneously at $W\approx 4t$.
However, $Q_{xy}\approx 0.5$ is still quantized in this case.
Therefore, it is appropriate to deduce that the edge band still has the non-trivial mass domain walls to host the corner states. Meanwhile, the mass-domain walls induced by edge gaps are filled with localized states.
At present, it is difficult to clearly demonstrate whether these edge gaps are closed or not due to the localized states.
A careful characterization of edge states under strong disorder requires further investigation.

\section{HOTI with edge-corner correspondence}\label{section4}

This section focuses on HOTIs with edge-corner correspondence, where the bulk-corner correspondence breaks down.
In particular, we pay attention to the weak disorder strength cases so that the SCBA can describe the disorder-induced band renormalization.

\subsection{edge band gap closing}

\begin{figure}[b]
   \centering
    \includegraphics[width=0.49\textwidth]{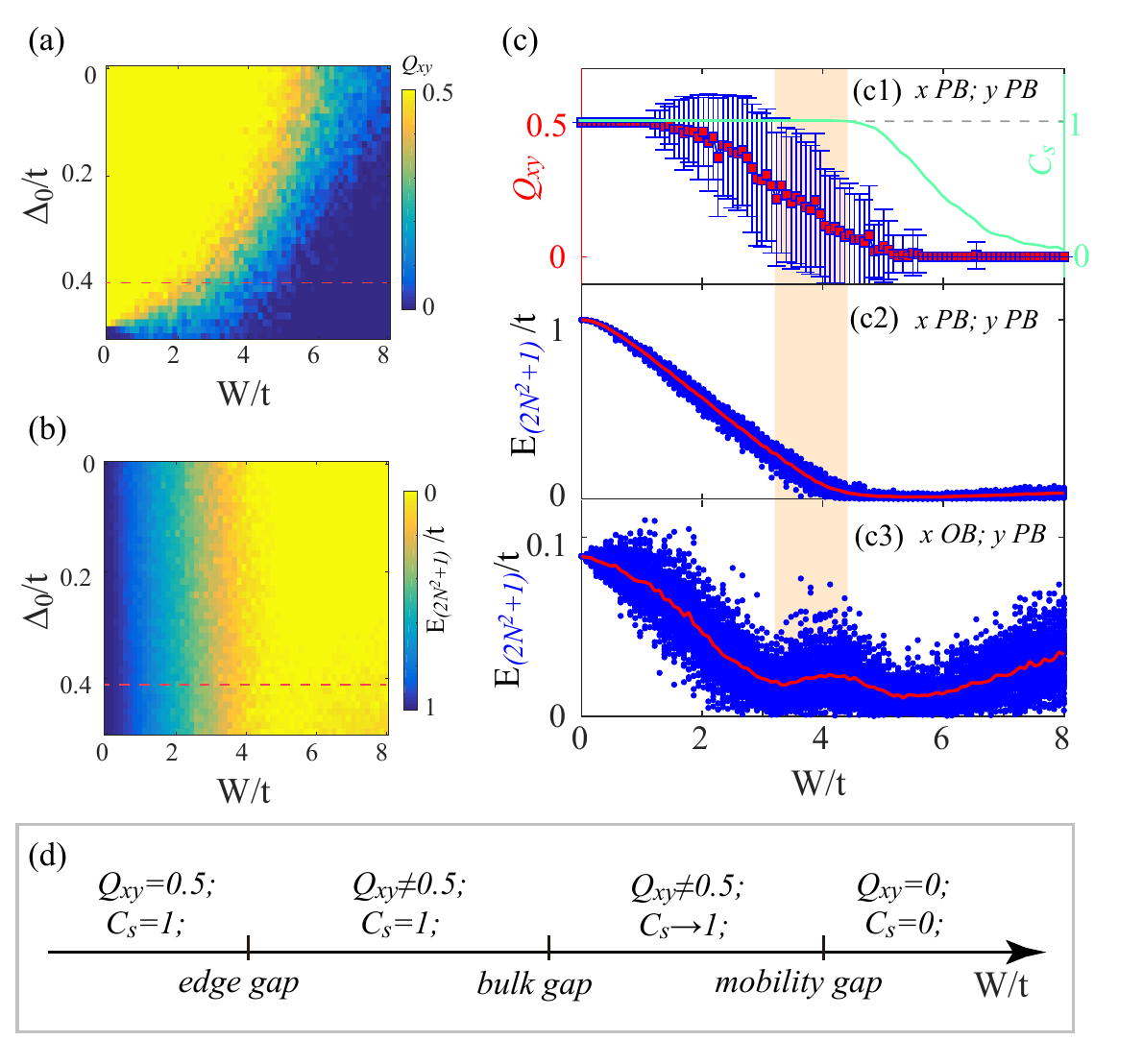}
    \caption{(Color online). (a) and (b) $Q_{xy}$ and the bulk energy gap versus $\Delta_0$ and disorder strength $W$. The dashed lines are $\Delta_0=0.4t$. (c) shows the typical plot with $\Delta_0=0.4t$.  (c1)-(c3) $C_s$, $Q_{xy}$, bulk gap and edge band gap evolution with the increasing of $W$. The parameters are  $m=t$ and $\Delta_x=0.5t$. The sample size is $N=40$.  (d) A schematic plot of gap evolution by increasing disorder strength. }
   \label{f6}
\end{figure}

Next, we show that the bulk-corner correspondence breaks down for the Hamiltonian in Eq.~(\ref{EQ1}) with $\Delta_0\neq 0$.
We plot the $Q_{xy}$ and the bulk energy gap as a function of $W$ and $\Delta_0$ in Fig.~\ref{f6}(a) and (b), respectively.
The quantized quadrupole moment $Q_{xy}=0.5$ (the HOTIs) region decreases with $\Delta_0$ increasing in Fig.~\ref{f6}(a), while the bulk energy gap is independent of $\Delta_0$ as shown in Fig.~\ref{f6}(b).
This is inconsistent with the fact that the bulk energy gap of clean samples is independent of $\Delta_0$ for such cases. Therefore, the bulk-corner correspondence no longer holds.

To gain further insight into the edge-corner correspondence of the system, we compare the spin Chern number $\mathcal{C}_s$ with the quadrupole moment $Q_{xy}$ and plot them against $W$ with fixed $\Delta_0=0.4t$ as shown in Fig.~\ref{f6}(c1).
Notably, the spin Chern number $\mathcal{C}_s\approx1$ is more robust than the quadrupole moment $Q_{xy}=0.5$, and it remains a quantized value for $W<5t$.
These results are distinct from the bulk-corner correspondence cases where $C_s$ and $Q_{xy}$ curves overlap as shown in Figs.~\ref{f3}(a) and (b).
Furthermore, we show the bulk and edge energy gaps in Figs.~\ref{f6}(c1) and (c2), to compare $\mathcal{C}_s$ and $Q_{xy}$ with the two energy gaps.
It is clear that the robustness of $Q_{xy}$ is mainly determined by the edge band gap closing at $W\approx3t$,
while $\mathcal{C}_s$ remains quantized until the bulk energy gap closes at $W\approx4.5t$.
This justifies the fact that $Q_{xy}=0.5$ results from the mass domain walls of edge modes while $\mathcal{C}_s$ is protected by the bulk gap. 

Now we come to summarize the gap closing process for the edge-corner correspondence case in Fig.~\ref{f6}(d).
In the clean limit, the system is topological non-trivial, with $\mathcal{C}_s=1$ and $Q_{xy}=0.5$ that
lead to helical edge modes in the bulk gap and corn states in the edge gap.
By turning on $W$, the edge gap fills up and closes, destroying the quantized quadrupole moment, i.e., $Q_{xy}\neq0.5$.
Further increasing $W$, the bulk energy gap is closed while $\mathcal{C}_s$ is still quantized because of the mobility bulk gap. At last, both $\mathcal{C}_s$ and $Q_{xy}$ equal to zero, and the system is an Anderson insulator when the mobility gap is closed.
We note that, because the bulk and edges gaps close simultaneously in Sec.~\ref{section3} with $\Delta_0=0$,
$\mathcal{C}_s$ and $Q_{xy}$ curves overlap, and the system recovers the bulk-corner correspondence.

\begin{figure}[t]
   \centering
    \includegraphics[width=0.49\textwidth]{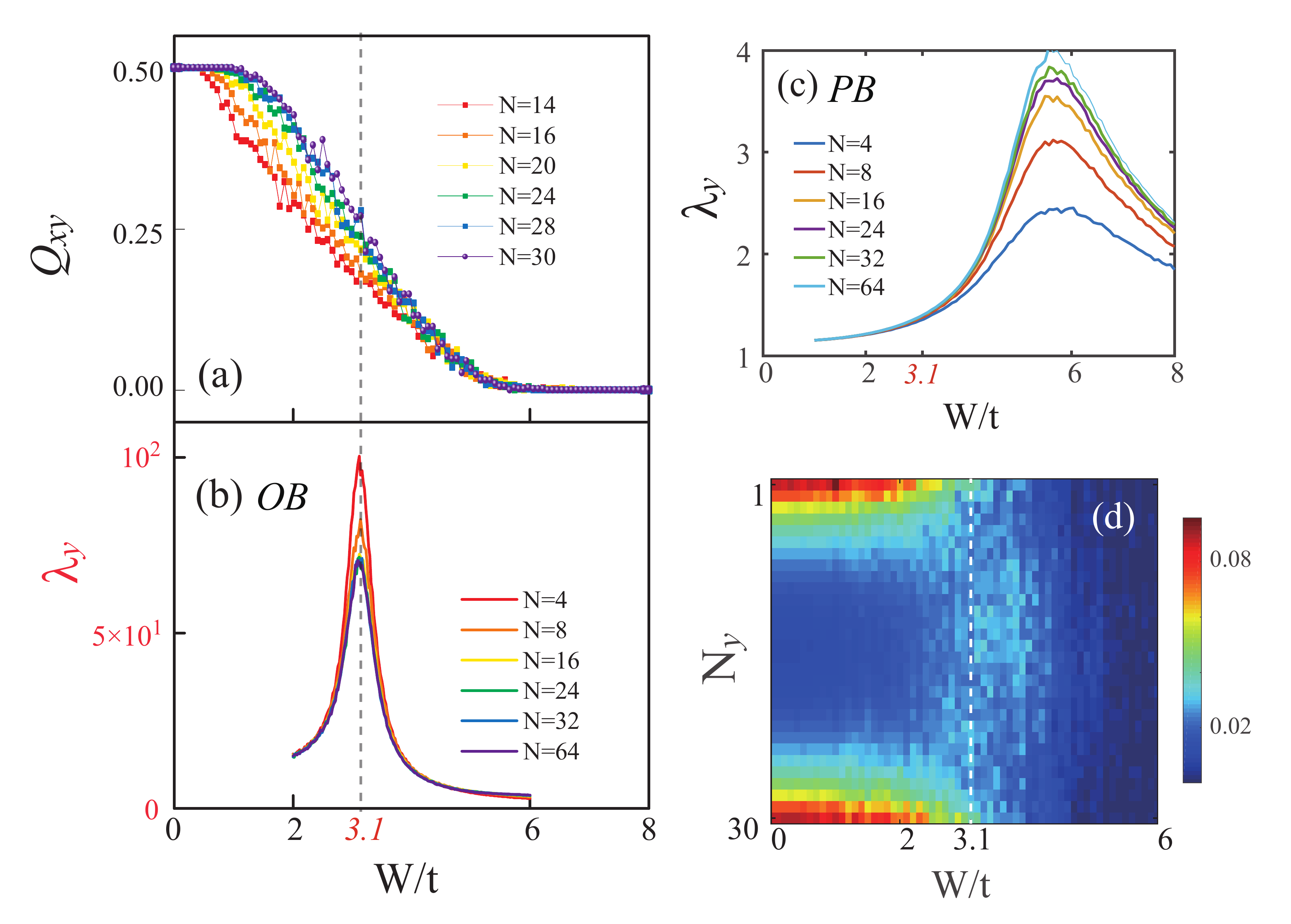}
    \caption{(Color online). (a) shows $Q_{xy}$ versus disorder strength $W$. (b) and (c) show  $\lambda_y$ versus $W$ for open boundary and periodic boundary conditions, respectively. (d) is the typical plot of the averaged wavefunction $|\psi(i_x=1,i_y)|^2$ with $i_y\in [1,N]$ for different disorder strength. The open boundary condition is adopted with $\Delta_0=0.4t$, $\Delta_x=0.5t$, $m=t$. }
   \label{f7}
\end{figure}

At last, we evaluate the localization length $\lambda_y$ along $y$ direction with periodic and open boundary conditions to accurately determine the edge and bulk energy gap closing points, respectively.
In Fig. \ref{f7}(b), the localization length has a peak sitting at $W\approx 3.1t$ under the open boundary condition. Besides, the position is independent of the sample size. On the other hand, such a peak disappears in the periodic boundary cases [see Fig. \ref{f7}(c)], which implies that the peak shown in Fig. \ref{f7}(b) should be related to the features of edge states\cite{HOTIq1}.
 Next, we compare $\lambda_y$ with the scaling of $Q_{xy}$, as shown in Fig.~\ref{f7}(a).
 $Q_{xy}$ gradually loses quantization near the peak at $W\approx3.1t$, where $Q_{xy}$ almost decreases to half of its quantized value. Moreover, we plot the disorder-averaged wave functions of corner states along $y$ direction $|\psi(i_x=1,i_y)|^2$ against $W$ with $i_y\in[1,N]$ explicitly.
 Figure \ref{f7}(c) clearly shows that the corner states spread into the edge along direction $y$ and disappear
when $W\approx 3.1t$.
Thus, we conclude that the sharp peak of $\lambda_y$ can be used to accurately determine the edge band gap closing point where the corner states disappear due to the edge-corner correspondence.

\subsection{gap closing based on SCBA}

\begin{figure}[b]
   \centering
    \includegraphics[width=0.49\textwidth]{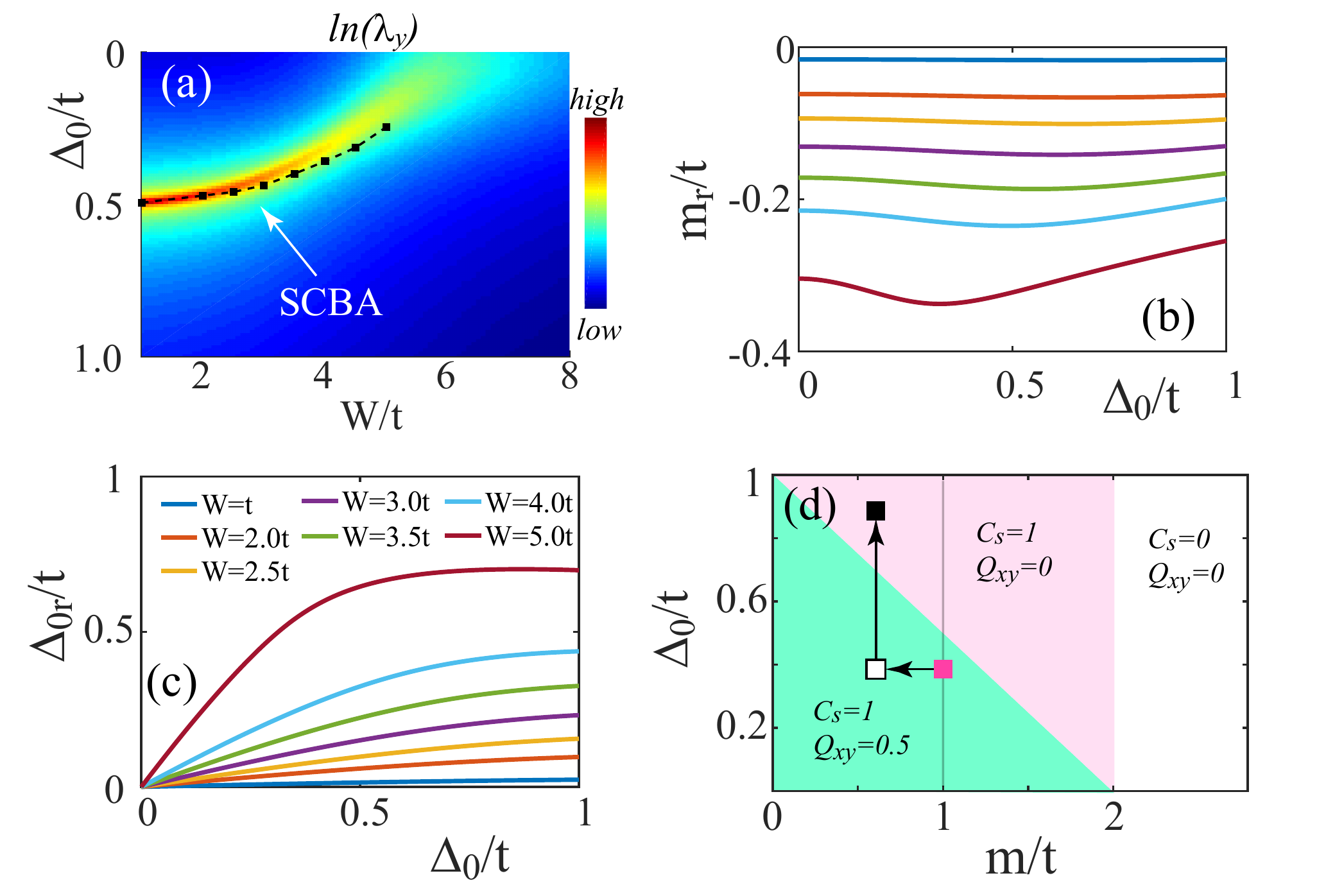}
    \caption{(Color online). (a) $\ln(\lambda_y)$ versus $(W,\Delta_0)$. $\lambda_y$ is the localization length along $y$ direction. The square and dashed line are obtained by SCBA. The sample size is $N=8$. (b) and (c) renomalized $m_r$ and $\Delta_{0r}$ versus $\Delta_0$ for different disorder strength $W$, respectively. (d) The schematic diagram of the path in the phase diagram. Parameters are the same as those in Fig. \ref{f7}. }
   \label{f8}
\end{figure}

To elucidate the edge band gap closing under disorder, we evaluate the disorder-induced band renormalization using the SCBA method.
At first glance, it seems to contradict the previous theory of disorder-induced topological Anderson insulators\cite{HOTIq1,HOTIq2} since gap closing induces a phase transition from $Q_{xy}=0.5$ (non-trivial) to $0$ (trivial). In the following, we will clarify it by showing that the renormalization to both topological mass $m$ and $\Delta_0$ term can give rise to a much richer phase diagram.

Generally, the topological invariants $C_s$ and $Q_{xy}$ of the Hamiltonian $\mathcal{H}({\bf k})$ are determined by $m$, $\Delta_0$ and $\Delta_{x,y}$. In particular, the edge gap in $x$ ($y$) direction is determined by a function $\bar{\Delta}_{x,y}=f(m,\Delta_0,\Delta_{x,y})$.
When the gap $\bar{\Delta}_{x}$ and $\bar{\Delta}_{y}$ have opposite sign, there are corner states near the domain walls, giving rise to $Q_{xy}=0.5$.
Therefore, we focus on the renormalization of $\Delta_0$ and $m$ in the following.
In the presence of disorder,
the band structure is normalized by the disorder-induced self-energy $\Sigma^r$ according to the SCBA.
$\Sigma^r$ can be decomposed as: $\Sigma^r=\sum_{ij}\Sigma^r_{ij}\tau_i \sigma_j $ with $i,j=0,x,y,z$.
The mass $m$ and $\Delta_{0}$ are renormalized into $m_R\equiv m + m_r =m - \Sigma^r_{zz} $ and $\Delta_{0R}\equiv \Delta_{0} + \Delta_{0r} =\Delta_{0} + \Sigma^r_{0x} $, respectively.
In Fig.~\ref{f8}(a), the phase boundary predicted by SCBA is perfectly fitted to gap-closing points indicated by the maximum localization length. This suggests that the phase transition from non-trivial to trivial can be described by the disorder-induced self-energy $\Sigma^r$.
It is important to notice that
the disorder-induced renormalization suggests $m_r<0$ and $\Delta_{0r}>0$ for various $\Delta_0$ when $\Delta_0\neq0$, shown in Figs.~\ref{f8}(b) and (c).

Now we can explain the disorder-induced phase transition from $Q_{xy}=0.5$ to $Q_{xy}=0$.
Figure~\ref{f8}(d) shows the phase diagram in the clean limit with phase boundaries determined by the edge band gap closing and spin Chern number [see appendix for more details].
If we start from the red square (with $\Delta_0=0.4t$ and $m=t$) in Fig.~\ref{f8}(d),
$m$ is renormalized to $m_R<m$, i.e., the red square shifts to the white square, and the sample tends to be more topological. It is consistent with the previous studies of disorder-induced topological Anderson insulators. However, since $\Delta_{0}\neq 0$, $\Delta_0$ is also renormalized along the vertical direction so that $\Delta_{0R}$ increases.
In addition, $\Delta_{0r}$ is larger than $m_r$ when $W\rightarrow5t$, as shown in Fig. \ref{f8}(b) and (c). The white square continuously shifts to the black square by crossing the phase boundary when the disorder is strong enough. Therefore, a phase transition from $Q_{xy}=0.5$ to $Q_{xy}=0$ happens with $\mathcal{C}_s=1$ unchanged. These predictions are consistent with the results shown in Fig. \ref{f6}(c). Furthermore, $m_r$ is almost independent of $\Delta_0$, while $\Delta_{0r}$ increases with an increase of $\Delta_0$. Thus, the critical disorder strength between $Q_{xy}=0.5$ and $Q_{xy}=0$ decreases for larger $\Delta_0$. This explains what we observe in Fig.~\ref{f6}(a).

\begin{figure}[t]
   \centering
    \includegraphics[width=0.43\textwidth]{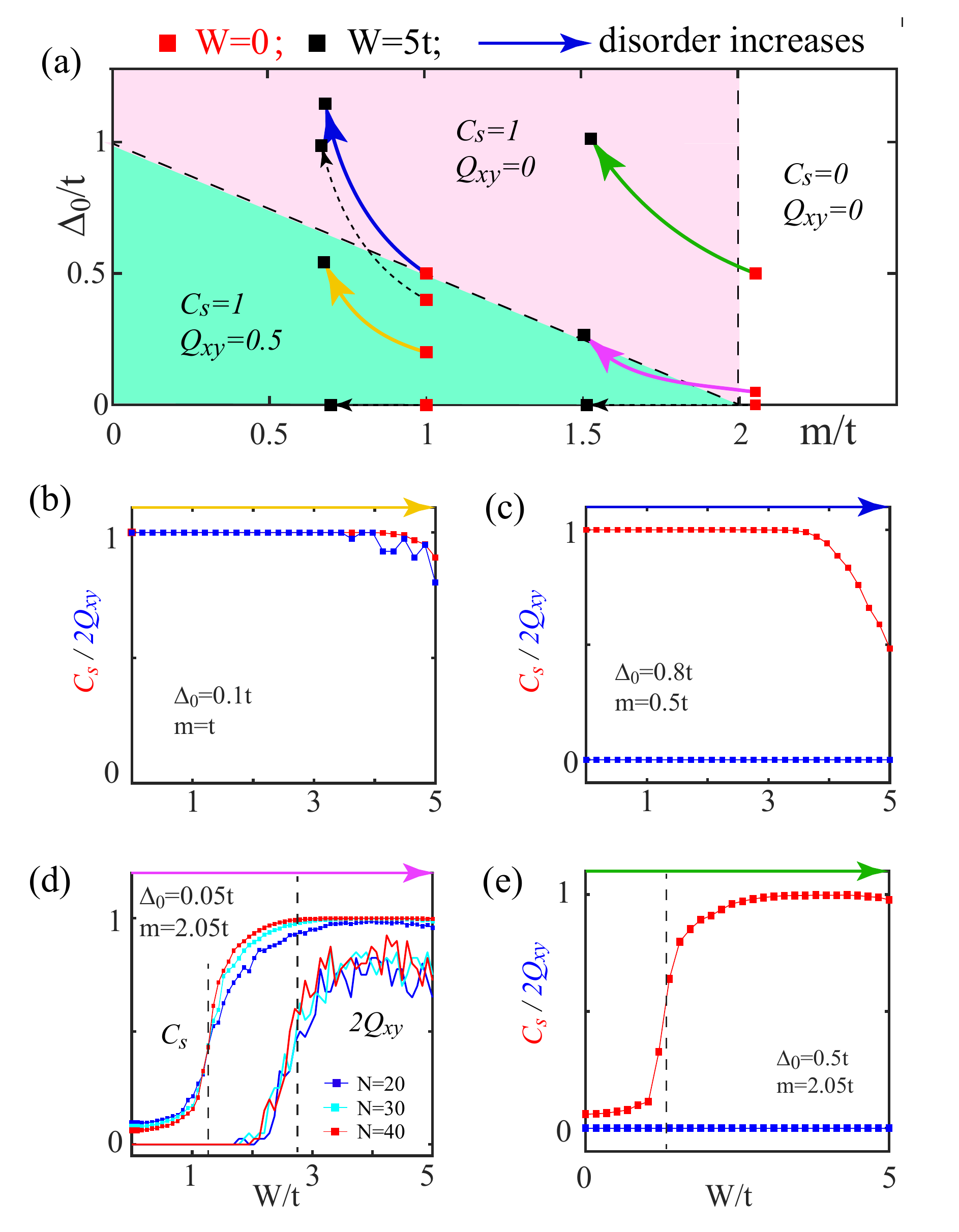}
    \caption{(Color online). (a) The flow under different parameters with disorder strength $W\in[0,5t]$. The black (red) squares correspond to $W=0$ ($W=5t$). The direction of the arrows suggests that disorder strength $W$ increases. All the curves are obtained based on the SCBA calculations. The arrows with dashed line are the three conditions already known. Four solid lines are four possible gap closing cases shown in (b)-(e).  Parameter is set as $\Delta_x=0.5t$. The phase boundary is $m=2t-2\Delta_0$. The sample size is $N=30$. }
   \label{f9}
\end{figure}

In the following, let us show that the renormalization to both $m$ and $\Delta_0$ can give rise to much richer phase transitions than the previously discovered TAI.
As shown in Fig~.\ref{f9}(a), we plot some possible paths in ($m$, $\Delta_0$) space. Similar to Fig.~\ref{f8}(d), the red square and the black square correspond to the starting point ($W=0$) and the final point (W=5t), respectively. The solid and dashed lines connecting the red and black squares are paths obtained by SCBA. The arrow suggests the renormalization direction as the disorder strength $W$ increases.

Let us verify the three dashed lines (that starting from $\Delta_0=0$ and $\Delta_0=0.4$) in Fig.~\ref{f9}(a), which are the three paths studied in previous sections. For $\Delta_0=0$, two different renormalization paths are given for $m=t$ and $m=2.05t$. They are correlated with
Figs.~\ref{f3}(a) and (b) as follow. $Q_{xy}=0.5$ and $C_s=1$ are unchanged for Fig. \ref{f3}(a), while the trivial one ($Q_{xy}=0$ and $C_s=0$) is transformed into nontrivial with $Q_{xy}=0.5$ and $C_s=1$ for Fig. \ref{f3}(a). These two paths are traditional since $\Delta_{0}=0$, and only $m$ is renormalized.
On the other hand, for the path starting from $\Delta_0=0.4t$ and $m=t$ as discussed in Fig. \ref{f9}(a),
$Q_{xy}=0.5$ decreases to zero with $C_s=1$  which corresponds to the case in Fig. \ref{f6}(c)

In addition, four new paths marked with the solid lines are shown in Fig. \ref{f9}(a).
First, the yellow path suggests that no phase transition exists, in agreement with the plot of $C_s$ and $Q_{xy}$ in Fig. \ref{f9}(b). Then, even when the starting point moves close to the phase boundary [see solid blue line in Fig. \ref{f9}(a)] the HOTAI is not available for larger $\Delta_0$ as verified in Fig. \ref{f9}(c).
Nevertheless, figure \ref{f9}(d) shows that rich phase transitions are possible as indicated by the magenta line in Fig. \ref{f9}(a).
The spin Chern number is changed $\mathcal{C}_s=0\rightarrow1$ with the renormalized topological mass $m_R\approx2t$ and then the edge band gap is closed with $Q_{xy}\approx 0.5$ [see appendix for more details].
At last, we find only $\mathcal{C}_s=0\rightarrow1$ for the case shown in Fig. \ref{f9}(e).

\section{discussion and summary}\label{section5}

In section \ref{section3}, we have discussed the existence of mass-domain-walls induced by edge gaps for bulk-corner correspondence case with $\Delta_0=0$, since $Q_{xy}\rightarrow0.5$ for the strong disorder strength. However, whether such edge gaps are closed or not is still unclear because of the existence of the localized bulk states.
Moreover, in section \ref{section4}, when the bulk-corner correspondence breaks down by disorder for $\Delta_0\neq0$, the edge band gap is closed and the mobility gap is still alive.  Due to the specific characteristics of HOTIs without bulk-corner correspondence, the gap closing of HOTIs under strong disorder could be more complicated. These problems deserve further investigation.

\begin{figure*}[t]
   \centering
    \includegraphics[width=0.8\textwidth]{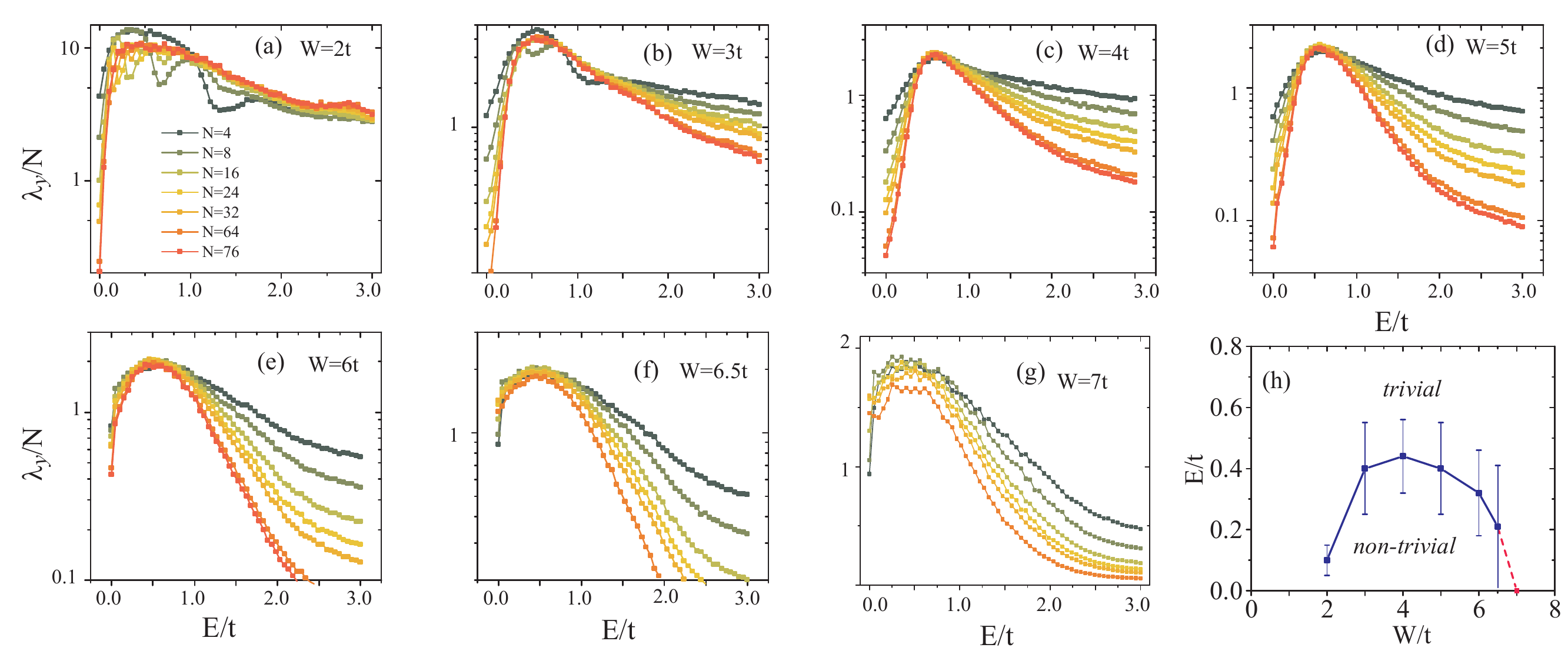}
    \caption{(Color online). The normalized localization length $\lambda_y/N$ versus energy $E$ for different sample sizes $N$. The disorder strength is: (a) $W=2t$, (b) $W=3t$, (c) $W=4t$, (d) $W=5t$, (e) $W=6t$, (f) $W=6.5t$, (g) $W=7t$, respectively. (h) shows the summation of the critical points.
     Parameters are set as $m=2.05t$, $\Delta_0=0$, and $\Delta_x=0.5t$. The periodic boundary condition is adopted.}
   \label{f10}
\end{figure*}

In summary, we studied the gap closing for HOTIs with or without bulk-corner correspondence. For the bulk-corner correspondence case, we found that the bulk-corner correspondence still holds after considering the disorder effect. The existence of the mobility gap was also revealed. Furthermore, we demonstrated that the mobility gap closing induces the phase transition from HOTIs to Anderson insulator, and the quadrupole moment $Q_{xy}$ approaches zero under strong disorder correspondingly. We concentrated on the weak disorder strength regions for the edge-corner correspondence case, where the disorder-induced band renormalization dominates. Importantly, the disorder-induced phase transition from $Q_{xy}=0.5$ to $Q_{xy}=0$ was obtained. This transition originates from the disorder-induced renormalization of both $m$ and $\Delta_0$, distinct from the HOTAI cases. We found several phase transitions based on the renormalization, and the corresponding gap closing was also clarified.

\section{ACKNOWLEDGEMENT}

We are grateful to Qiang Wei, Hongfang Liu, and Hua Jiang for fruitful discussions. This work was supported by NSFC under Grant N0. 11974256, and the NSF of Jiangsu Province under Grant N0. BK20190813. Z.Q.Z. was supported by National Basic Research Program of China (Grant No. 2019YFA0308403), NSFC under Grant No. 11822407, and a Project Funded by the Priority Academic Program Development of Jiangsu Higher Education Institutions.

Y.S.H. and Y.R.D. contribute equally to this work.

\appendix
\section{the mobility gap for HOTAI}\label{appendixA}

We present the mobility gap evolution of the HOTAI shown in Fig. \ref{f3}(b) with $m=2.05t$. The main results are similar to those in Fig. \ref{f4}. The mobility gap first increases, shown in Fig. \ref{f10}(h). Then, the mobility gap decreases due to the localization effect.

\section{The bulk-corner, edge-corner, and bulk-edge correspondence}\label{appendixB}

\begin{figure}[t]
   \centering
    \includegraphics[width=0.45\textwidth]{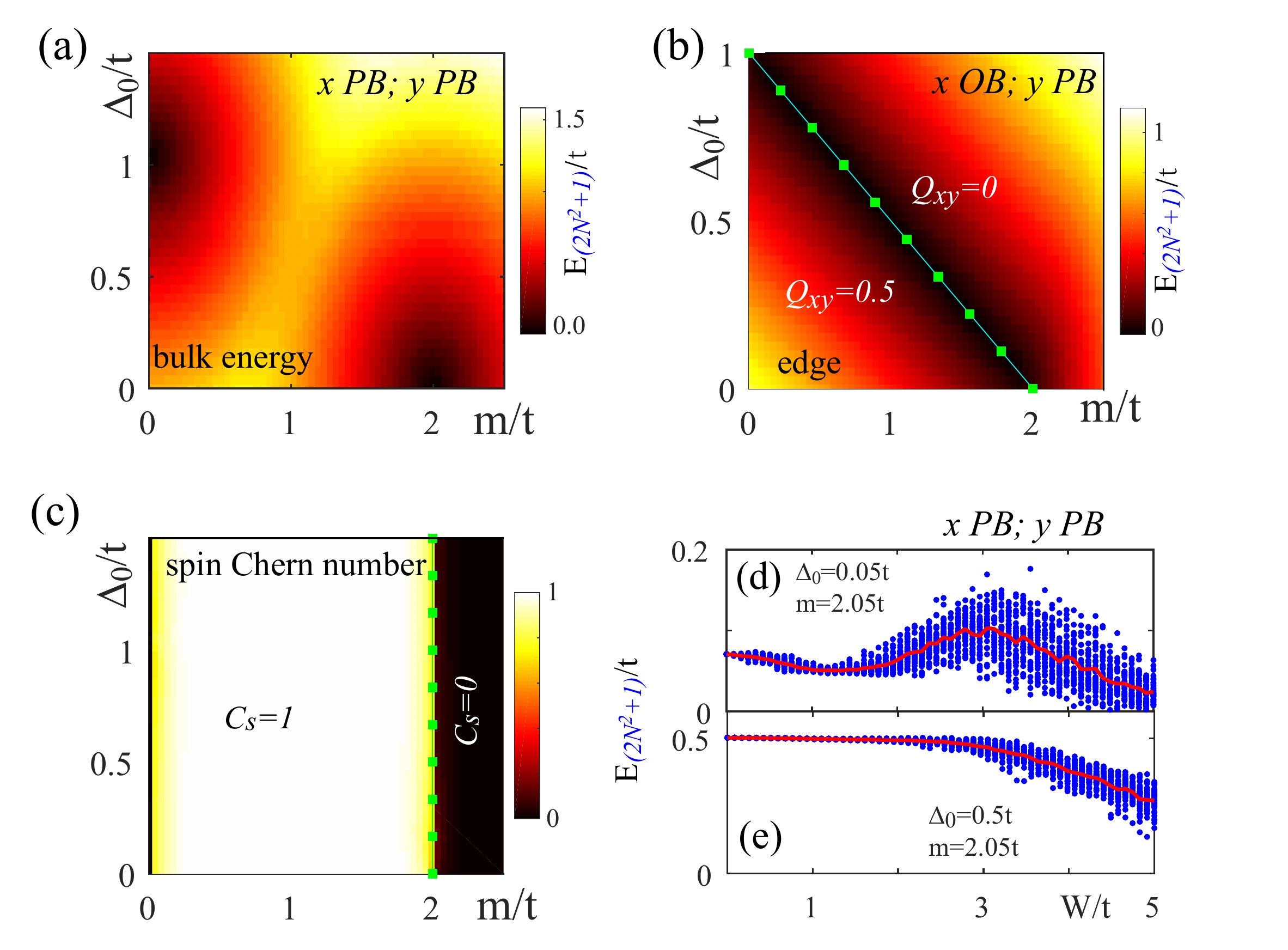}
    \caption{(Color online). (a) and (b) The bulk energy gap and edge band gap versus $m$ and $\Delta_0$, which are calculated under periodic and open boundary along $x$ direction, respectively. Periodic boundary along $y$ direction is used. (c) The evolution of spin Chern number. The green solid lines with square marks are the phase boundary. The quadrupole moment with $Q_{xy}=0$ and $Q_{xy}=0.5$ are separated by $m=2t-2\Delta_0$ (d) and (e) the evolution of bulk energy gap versus disorder strength $W$ for Fig. \ref{f9}(d) and (e), respectively. $\Delta_x$ is fixed at $0.5t$. Other parameters are given in the figure. The sample size is $N=20$.}
   \label{f11}
\end{figure}

For the model shown in Eq. (\ref{EQ1}), there are three possible cases for the bulk-boundary correspondence.
(i) When bulk-corner correspondence holds, the bulk energy (mobility) and edge band gap should be closed simultaneously with $Q_{xy}$ varying.
(ii) The bulk-edge correspondence in this paper means that $\mathcal{C}_s=1$  ensures the existence of edge states with or without edge band gaps,
where  $\mathcal{C}_s$ changes only when the bulk energy (mobility) gap is closed.
(iii) For edge-corner correspondence, $Q_{xy}$ that determines the corner states is correlated with the edge band gap instead of bulk energy gap.

To clarify the bulk-corner, bulk-edge, and edge-corner correspondence\cite{HOTIbe1}, we plot the evolution of bulk energy gap and edge band gap in Figs. \ref{f11}(a) and (b) for samples without disorder. Fig. \ref{f11}(c) shows the spin Chern number $\mathcal{C}_s$.
  For $\Delta_0=0$, the bulk gap is closed when $m=2t$. Furthermore, $\mathcal{C}_s$ and $Q_{xy}$ change simultaneously, where the bulk-edge and bulk-corner correspondence exist.
When $\Delta_0\neq0$, the bulk energy gap is closed only when $(m=0,\Delta_0=0)$ and $(m=0,\Delta_0=t)$ [see Fig. \ref{f11}(a)]. For the rest regions, the bulk gap always exists. However, $\mathcal{C}_s$ changes only when $m=2t$, which is independent of $\Delta_0$ [see Fig. \ref{f11}(c)]. Thus, the bulk-edge correspondence breaks down.
By comparing the bulk energy gap in Fig. \ref{f11}(a) with $Q_{xy}$ shown in Fig. \ref{f11}(b), it is obvious that the bulk-corner correspondence breaks. Nevertheless, the edge-corner correspondence exists.
We note that, in Fig. \ref{f11}(b), the bulk energy gap is used to replace the edge energy gap for convenience when the edge states are absent.
Further, we compare the bulk energy gap in Figs. \ref{f11}(d) and (e) and $C_s$ shown in Fig. \ref{f9}(d) and (e) and find that the bulk energy gap is not closed when $C_s$ jumps from zero to one or vice versus when the bulk-edge correspondence breaks down.
The $C_s$  changes value only when $m=2t$, which is corresponding to the gap closing condition in BHZ model.



\end{document}